\documentclass[conference]{IEEEtran}

\usepackage[T1]{fontenc}
\usepackage[utf8]{inputenc}

\usepackage[kerning=true]{microtype}

\usepackage{pgfplots}
\pgfplotsset{compat=1.8}

\usepackage{color}

\usepackage[colorlinks]{hyperref}

\definecolor{mycitecolor}{rgb}{0.5,0,0}
\definecolor{mylinkcolor}{rgb}{0.0, 0.5, 1.0}
\definecolor{myurlcolor}{rgb}{0.0, 0.28, 0.67}

\definecolor{bleudefrance}{rgb}{0.19, 0.55, 0.91}
\definecolor{cardinal}{rgb}{0.77, 0.12, 0.23}

\hypersetup{citecolor=bleudefrance}
\hypersetup{linkcolor=cardinal}
\hypersetup{urlcolor=myurlcolor}

\usepackage{url}

\usepackage{amsmath,amssymb,amsfonts}
\usepackage{pifont}
\usepackage{graphicx}

\usepackage{cite}

\usepackage{algorithm}
\usepackage{algpseudocode}

\usepackage{bmpsize}
\usepackage{lipsum}

\usepackage{caption}
\usepackage{subcaption}
\usepackage{paralist}

\newcommand{\etal}{\textit{et al.}}

\definecolor{burgundy}{rgb}{0.5, 0.0, 0.13}
\definecolor{burntorange}{rgb}{0.8, 0.33, 0.0}
\definecolor{byzantine}{rgb}{0.74, 0.2, 0.64}
\definecolor{darkcyan}{rgb}{0.0, 0.55, 0.55}
\definecolor{amaranth}{rgb}{0.9, 0.17, 0.31}
\definecolor{darkturquoise}{rgb}{0.0, 0.81, 0.82}

\mathchardef\mhyphen="2D 
\newcommand{\cmark}{\ding{51}}%
\newcommand{\xmark}{\ding{55}}%

\renewcommand\vec{\mathbf}

\newcommand{\size}[1]{\lvert #1 \rvert} 

\newcommand{\prob}[1]{\Pr \left[ #1 \right]}

\newcommand{\ExpVal}[2]{\mathbb{E}_{#1}[#2]}

\newcommand{\ledger}{\mathcal{L}}

\newcommand{\our}{\textsc{Mitosis}}

\newcommand{\ToKGenerateProof}{\mathsf{ToK.GenerateProof}}
\newcommand{\ToKVerifyProof}{\mathsf{ToK.VerifyProof}}
\newcommand{\ToKtag}{\tau}
\newcommand{\ToKproof}{\pi}

\newcommand{\prover}{\mathcal{P}}
\newcommand{\verifier}{\mathcal{V}}

\newcommand{\ToALock}{\mathsf{ToA.Lock}}
\newcommand{\ToAprooflock}{\pi_{\mathrm{lock}}}
\newcommand{\ToAClaim}{\mathsf{ToA.Claim}}
\newcommand{\ToAproofclaim}{\pi_{\mathrm{claim}}}
\newcommand{\ToAproofabort}{\pi_{\mathrm{abort}}}
\newcommand{\ToAResolve}{\mathsf{ToA.Resolve}}

\def\BibTeX{{\rm B\kern-.05em{\sc i\kern-.025em b}\kern-.08em
    T\kern-.1667em\lower.7ex\hbox{E}\kern-.125emX}}

\begin{document}

\title{\our: Practically Scaling\\ Permissioned Blockchains}

\author{\IEEEauthorblockN{%
    Giorgia Azzurra Marson\IEEEauthorrefmark{1}, %
    Sebastien Andreina\IEEEauthorrefmark{2}, %
    Lorenzo Alluminio\IEEEauthorrefmark{3}, %
    Konstantin Munichev\IEEEauthorrefmark{4} and %
    Ghassan Karame\IEEEauthorrefmark{5}%
    }
\IEEEauthorblockA{NEC Laboratories Europe\\ Heidelberg, Germany\\
Email: \IEEEauthorrefmark{1}giorgia.marson@neclab.eu,
\IEEEauthorrefmark{2}sebastien.andreina@neclab.eu,
\IEEEauthorrefmark{3}lorenzo.alluminio@neclab.eu,\\
\IEEEauthorrefmark{4}konstantin.munichev@neclab.eu,
\IEEEauthorrefmark{5}ghassan@karame.org}%
}

\maketitle

\begin{abstract}
Scalability remains one of the biggest challenges to the adoption of permissioned blockchain technologies for large-scale deployments.
Namely, permissioned blockchains typically exhibit low latencies, compared to permissionless deployments---however at the cost of poor scalability.
As a remedy, various solutions were proposed to capture ``the best of both worlds'', targeting low latency and high scalability simultaneously. Among these, blockchain sharding emerges as the most prominent technique.
Most existing sharding proposals exploit features of the permissionless model and are therefore restricted to cryptocurrency applications. 
A few permissioned sharding proposals exist, however, they either make strong trust assumptions on the number of faulty nodes or rely on trusted hardware or assume a static participation model where all nodes are expected to be available all the time.
In practice, nodes may join and leave the system dynamically, which makes it challenging to establish how to shard and when.

In this work, we address this problem and present~\our, a novel approach to practically improve scalability of permissioned blockchains. Our system allows the dynamic creation of blockchains, as more participants join the system, to meet practical scalability requirements. Crucially, it enables the division of an existing blockchain (and its participants) into two---reminiscent of mitosis, the biological process of cell division. \our{} inherits the low latency of permissioned blockchains while preserving high throughput via parallel processing.
Newly created chains in our system are fully autonomous, can choose their own consensus protocol, and yet they can interact with each other to share information and assets---meeting high levels of interoperability.
We analyse the security of \our{} and evaluate experimentally the performance of our solution when instantiated over Hyperledger Fabric. Our results show that \our{} can be ported with little modifications and manageable overhead to existing permissioned blockchains, such as Hyperledger Fabric. As far as we are aware, \our{} emerges as the first workable and practical solution to scale existing permissioned blockchains. 
\end{abstract}

\section{Introduction}

Blockchains and decentralized applications thereof are evolving rapidly.
The initial wave of interest in cryptocurrencies, initiated with Bitcoin~\cite{bitcoin}, envisioned permissionless blockchains as an ideal solution to realize trustless payments over the Internet, allowing peers to exchange assets without the intermediation of financial institutions.
Despite the initial fame, Bitcoin and follow-up permissionless systems have been found to suffer a number of shortcomings~\cite{DBLP:conf/ccs/GervaisKWGRC16}, precluding their adoption for real-world applications.
A major obstacle to their widespread adoption is rooted in their probabilistic consistency and liveness guarantees, offering a rather weak notion of ``eventual consensus''. Concretely, although blocks are generated at a regular pace, the blockchain nodes cannot be certain that these blocks are stable in the ledger---they can only become more confident that a given block will not be reverted as more blocks are added ``on top'' of it.
Probabilistic finality of blocks directly reflects on the ledger in terms of transaction-confirmation time. This means that transactions cannot be confirmed with certainty, and after being included to the ledger, high-confidence confirmation is possible only once they are deep enough in the blockchain. Since confirmation time is slow, latency and throughput of permissionless systems are extremely limited compared to that of classical consensus protocols.
In contrast, permissioned blockchains provide \emph{finality}, meaning that once a block is included to the blockchain, it is already final (i.e., no rollback will be possible later on).
This makes permissioned blockchains an attractive, faster alternative to permissionless solutions, particularly for realistic deployments.
It is no surprise that prominent financial institutions are exploring permissioned blockchains to improve their services and modernize their businesses~\cite{DBLP:journals/corr/abs-2103-00254}, and legal aspects of cryptocurrencies are being discussed~\cite{DLTswitzerland}.
On the downside, permissioned-based consensus protocols scale rather poorly in the number of consensus nodes, which limits their deployment to small- and medium-scale scenarios.

Major efforts in the blockchain space have been devoted to improving scalability, with \emph{blockchain sharding} being the most prominent proposal. Blockchain sharding refers to the generic paradigm of employing multiple blockchains in parallel, the ``shards'', operating different and more lightweight instances of the same consensus protocol.
The idea is that running parallel instances allows boosting the transaction throughput roughly by a factor equal to the number of shards.
Notwithstanding the efficiency gain, popular solutions (such as Elastico~\cite{DBLP:conf/ccs/LuuNZBGS16}, Omniledger~\cite{DBLP:journals/iacr/Kokoris-KogiasJ17}, RapidChain~\cite{DBLP:conf/ccs/ZamaniM018}, and Monoxide~\cite{DBLP:conf/nsdi/WangW19}) can hardly meet both scalability and security requirements~\cite{DBLP:Avarikioti:etal:2020}.
In fact, most sharding proposals are designed for, and exploit features of, the permissionless model, which considerably restricts their practical suitability to cryptocurrency applications.
To the best of our knowledge, all sharding proposals assume a static participation model, meaning that all participating nodes (precisely, the correct ones) must be available all the time. This allows establishing upfront how many shards can be run in parallel, depending on the number of participants.
Given that participation in blockchain systems can change dynamically, we argue instead that new shards have to be created ad-hoc.
Finally, we note that existing sharding techniques assume homogenous blockchains running the same consensus protocol; for practical deployments, however, different shards might benefit from choosing their own consensus protocol independently of other shards.

In this work, we propose a novel approach for improving scalability of permissioned blockchains.
Inspired by the sharding paradigm and mindful of its limitations, we seek to leverage parallelism in a way that offers flexibility.
Namely, we envision a \emph{dynamic} blockchain ecosystem where new blockchains can be created as the need arises, and can evolve over time to meet specific scalability requirements.
In contrast to sharding, where all blockchain instances are highly coordinated and obey the same consensus, our solution lets the various blockchains in the system self-organize, in a decentralized manner, depending on their needs.

Our solution is inspired by mitosis, the cell division process in biology in which a parent cell divides into two or more child cells:
we propose a novel mechanism, dubbed \our, to practically give birth to new blockchains by splitting an existing one.
Based on this intuitive approach to create new blockchains, we investigate the necessary conditions so that the chain-splitting process can be operated securely.
Particularly, we analyse how trust assumptions on the parent chain impact the security of the child chains, and identify sufficient requirements on the parent chain, in terms of tolerated faults, to ensure consistency and liveness for both child chains despite crash- or Byzantine failures, thereby enabling to bootstrap trust from the parent chain to its child chains.

\our{} leverages parallelism in order to scale permissioned blockchains with dynamic membership (i.e., where new users can join at any time), as it allows increasing the number of system participants arbitrarily while preserving the high throughput of a small-scale system. Moreover, \our{} enables different block\-chains to communicate with each other, meeting high standards for blockchain interoperability. This means that users belonging to different blockchains can easily interact, e.g., to transfer assets from one blockchain to another.
Our design particularly supports knowledge transfer and asset transfer across blockchains~\cite{cryptoeprint:2019:1128}, thereby letting users prove statements about the state of their chain to users of different chains, as well as to move asset from one chain to another.
Our system also offers a chain-fusion procedure that, opposite to division, combines two blockchains into one  (as in the fusion of cells), in case some of the blockchains significantly shrink in size.

Our contributions can be summarized as follows:
\begin{itemize}
    \item We present~\our, a methodology to create new blockchains by recursively splitting an existing system into two child systems. We employ~\our{} to develop a flexible permissioned blockchain ecosystem for large-scale deployment, in which blockchains can self-organize to keep the system scalable and functional. Based on its increased dynamism, our solution improves upon sharding and is in fact compatible with existing schemes supporting sharding, by additionally allowing the dynamic creation of heterogeneous shards.
    \item We analyse the security requirements for blockchain splitting, particularly in terms of tolerated faults, and we discuss techniques to instantiate chain division such that the robustness requirements are met.
    \item We show how to integrate our proposal in Hyperledger Fabric which only supports homogenous shards (with no communication between shards).
\end{itemize}

As far as we are aware, \our{} is the first complete and practical blockchain system which securely enables the creation of new blockchains as the need arises, and allows them to evolve over time to meet specific scalability requirements.
We believe that such a model fits very well with the current deployments of permissioned blockchains, such as Hyperledger Fabric, as it enables the creation of a flexible, scalable, and secure system for permissioned blockchains.

The remainder of this paper is organized as follows.
In Section~\ref{sec:problem} we discuss the problem statement in detail.
We present our solution in Section~\ref{sec:solution}, and analyse its security in Section~\ref{sec:analysis}.
In Section~\ref{sec:fabric} we discuss a practical instantiation of our blockchain ecosystem based on Hyperledger Fabric.
We discuss implementation details in Section~\ref{sec:evaluation}, and we report on the performance of our~\our{} instantiation based on an empirical evaluation.
We conclude the paper in Section~\ref{sec:conclusion}. 
\section{Problem Statement \& Background}
\label{sec:problem}

\begin{table*}[tb]
    \centering
    \caption{Evaluation of existing sharding solutions and our proposal (cf.~Section~\ref{sec:solution}) according to the criteria:
    Participation Model,
    Transaction Model,
    Support for Heterogeneous Shards;
    Support for Dynamic Sharding;
    Support for Shard Fusion.
    }
    \begin{tabular}{|l||c|c|c|c|c|}
     \hline
     \textbf{Protocol} & \textbf{Participation Model} & \textbf{TX Model} & \textbf{Heterogeneous Shards} & \textbf{Dynamic Sharding} & \textbf{Shard Fusion}\\ [0.5ex]
     \hline
     OmniLedger~\cite{DBLP:journals/iacr/Kokoris-KogiasJ17} & Permissionless & UTXO & \xmark & \xmark & \xmark \\
     RapidChain~\cite{DBLP:conf/ccs/ZamaniM018} & Permissionless & UTXO & \xmark & \xmark & \xmark \\
     Monoxide~\cite{DBLP:conf/nsdi/WangW19}  & Permissionless & UTXO & \xmark & \xmark & \xmark \\
     Horizontal channels~\cite{DBLP:conf/esorics/AndroulakiCCK18}& Permissioned & UTXO & \xmark & \xmark & \xmark\\
     AHL~\cite{DBLP:conf/sigmod/DangDLCLO19}& Permissioned & account-based & \xmark & \xmark & \xmark\\
     SharPer~\cite{DBLP:conf/sigmod/AmiriAA21}& Permissioned & account-based & \xmark & \xmark & \xmark\\
     \hline
    \textbf{\our{} (this work)} & Permissioned & account-based & \cmark & \cmark & \cmark\\
    \hline
    \end{tabular}
\end{table*}

In this section, we discuss in detail the problem addressed by our work.
Along the way, we also introduce background concepts and terminology.

\subsection{Distributed Consensus}

A blockchain protocol allows users to agree on a totally ordered sequence of transactions, i.e., a decentralized transaction ledger, to enable a consistent execution of these transactions in a distributed system.
Blockchains are therefore instantiations of total-order broadcast (a.k.a.~consensus) protocols.
Their main properties are expressed in terms of \emph{consistency}, meaning that the various participants agree on the ledger state, and \emph{liveness}, ensuring that transactions are included to the local ledgers of participants relatively quickly.
Distributed consensus protocols are designed to be resilient to a limited number of failures, thereby tolerating crashes and Byzantine faults respectively.%
\footnote{A Byzantine participant may deviate from the prescribed protocol in arbitrary ways, and even be controlled by an attacker.}
Namely, crash-fault and Byzantine-fault tolerance (CFT, resp.~BFT) require consistency and liveness to hold despite some of the participants being faulty.

Depending on the participation model, blockchains can be categorized into \emph{permissionless}, run among anonymous and mutually untrusted participants, and \emph{permissioned}, where users have explicit identities known to everybody at the protocol outset.
Permissionless blockchains provide a relatively weak consistency property: a transaction is more likely to be stable the deeper it is the ledger. Such probabilistic guarantee implies a slow confirmation time---about~10 minutes for Bitcoin and~5 minutes for Ethereum---which severely limits throughput to at most tens of transactions per second~(tps) for Bitcoin and Ethereum. 
In contrast, permissioned blockchains can use classical consensus protocols, which offer finality and therefore provide a much lower latency---popular consensus implementations can confirm thousands of transactions per second~\cite{FastBFT}.
On the downside, consensus protocols require a few rounds of interaction among all participants, requiring high communication complexity to reach agreement on each block entry (typically~$\mathcal{O}(n^2)$ where~$n$ is the number of participants), which severely hinders scalability. For instance, increasing the number of consensus nodes from~50 to~100 reduces the throughput from around~1000 to~100~tps.

\subsection{Challenges in Scaling Permissioned Blockchains}

Scalability remains the major challenge for the adoption of permissioned blockchains in real-world applications. Most existing solutions, e.g., FastBFT~\cite{FastBFT} and Hotstuff~\cite{Hotstuff}, aim at reducing the communication complexity.
Despite improving performance, all these solutions still rely on a classical consensus algorithm at their core, and the effective scalability gain is limited to one order of magnitude at best (from a few tens to hundreds of nodes).
Other scalability proposals fall in the domain of permissionless systems. Broadly, these proposals provide ``on-chain''  (or ``layer~1'') solutions such as sharding and DAG-based protocols~\cite{DBLP:journals/iacr/SompolinskyZ13,DBLP:journals/iacr/SompolinskyLZ16,tangle}, that directly operate on the consensus layer, and ``off-chain'' (or ``layer~2'') solutions, e.g., payment channels~\cite{bitcoinlightning,raiden} and side-chains~\cite{peggedsidechains,BCC:LSFK2017}, which handle the smart-contract layer only.
In this work, we focus on Layer~1 solutions.

Sharding appears as the most promising on-chain method to improve scalability and performance of blockchain protocols, with prominent instantiations such as RapidChain~\cite{DBLP:conf/ccs/ZamaniM018} and Monoxide~\cite{DBLP:conf/nsdi/WangW19}.
These solutions are designed for the permissionless model, particularly, they assume a UTXO model which does not generalize beyond cryptocurrency applications.
In the context of permissioned blockchains, Androulaki~\etal~\cite{DBLP:conf/esorics/AndroulakiCCK18} propose horizontal channels envisioned for Hyperledger Fabric. This proposal however also relies on the UTXO model, for enabling fast cross-shard transactions, hence its applicability to Fabric and other permissioned systems is unclear.  Besides, it implements sharding at the smart-contract layer and not on the consensus layer.

A sharding proposal built on Fabric is Attested HyperLedger (AHL)~\cite{DBLP:conf/sigmod/DangDLCLO19}. In AHL, each shard runs an optimized consensus protocol based on PBFT~\cite{DBLP:conf/osdi/CastroL99}, requiring consensus nodes to run trusted hardware to prevent Byzantine nodes from equivocating, thereby reducing the tolerated faults from~$\frac{n-1}{3}$ to $\frac{n-1}{2}$ (this is similar to FastBFT~\cite{FastBFT}).
In the same vein, SharPer~\cite{DBLP:conf/sigmod/AmiriAA21} aims at improving scalability of sharded permissioned blockchains, however, it dramatically limits the number of tolerated faults (e.g., $f \ll \frac{n}{3}$ for Byzantine faults)  and leverages this assumption to deterministically create~$\frac{n}{3f+1}$ shards that provably meet the consensus bound (this is analogous to the deterministic assignment scheme we discuss in Section~\ref{sec:division:validator:assignment}).
The only solution we are aware of that proposes a dynamic sharding approach is GearBox, a concurrent and independent work by David~\etal~\cite{gearbox} that leverages the safety-liveness dichotomy to decrease the shard size while preserving security. GearBox uses a control chain (which is assumed to be always live) to monitor the progress of the other shards, and it triggers shard reconfigurations dynamically whenever a deadlock is detected.

All the aforementioned solutions follow a common theme: they assume a large, fixed set of nodes and make it scale via sharding. This entails partitioning the set of nodes into~$m$ subsets---where $m$ is the number of shards---at the protocol onset, and parallelize transaction processing among the~$m$ shards, with the effect of boosting throughput roughly by a factor~$m$.
These solutions make the implicit assumption that all participating nodes are fixed upfront and keep participating in the consensus throughout the lifetime of the system.
Moreover, existing sharding solutions are rigid in enforcing all shards to be homogeneous, i.e., they run the same consensus protocol and ensure security under the same conditions. Some applications may however benefit from a more flexible sharding scheme that allows different shards to run different consensus protocols, so that each shard can choose the best option given local conditions.
This is particularly true under a dynamic-participation model, where new participants may join and existing participants may leave the system.
Various works in the permissionless model recognise that dynamic participation is desirable in practice. However, to the best of our knowledge, existing sharding solutions do not offer support for dynamic participation.
Under a dynamic participation model, it is not clear a priori how to shard, and when, in order to ensure optimal throughput. Existing sharding systems instead assume that these optimal conditions are known at initialization time.
We believe these limitations may challenge the adoption of sharding in practice.

Ideally, a truly scalable system should be able to dynamically adapt to external conditions, triggering sharding under high participation, and being able to merge shards in case of low participation.
To the best of our knowledge, currently there is no solution for permissioned blockchains that can reactively self-organize to meet optimal performance.

\section{\our: Overview and Design}
\label{sec:solution}

In this section we present~\our, our proposal to realize a secure, scalable, and flexible system of autonomous and interoperable blockchains.
Our goal is to design an effective approach to mitigate the scalability challenges in permissioned blockchains (cf. Section~\ref{sec:problem}). \our{} can be instantiated within existing permissioned blockchain frameworks with minimal modifications, as we discuss in Section~\ref{sec:fabric} and show empirically in Section~\ref{sec:analysis}.

\subsection{System Model and Assumptions}

We assume the standard blockchain communication model where users communicate with each other over a partially synchronous network.
We consider a permission-based model, where explicit registration is required for becoming a member of the system.
Conforming with most existing permissioned blockchains, users can have the following roles:
\emph{Clients}, or regular users, utilize the service provided by the blockchain. They submit requests in the form of transactions (e.g., a trading request in financial applications).
\emph{Validators}, or blockchain nodes, verify the clients' transactions and commit them to the blockchain, so that the corresponding requests are processed.
For the sake of abstraction, we assume a membership service maintaining members' information in a dedicated registry, so that members can retrieve information on-demand. The registry provides a means to identify members among each others, and acts as a discovery mechanism for new members.
Upon registering, a user~$u$ obtains an account~$A_u$ linking the user's identity~$u$, the corresponding public key~$\mathrm{pk}_u$, and possibly additional information about the user, depending on the application.

\begin{figure*}
\centering
\includegraphics[width=0.9\linewidth]{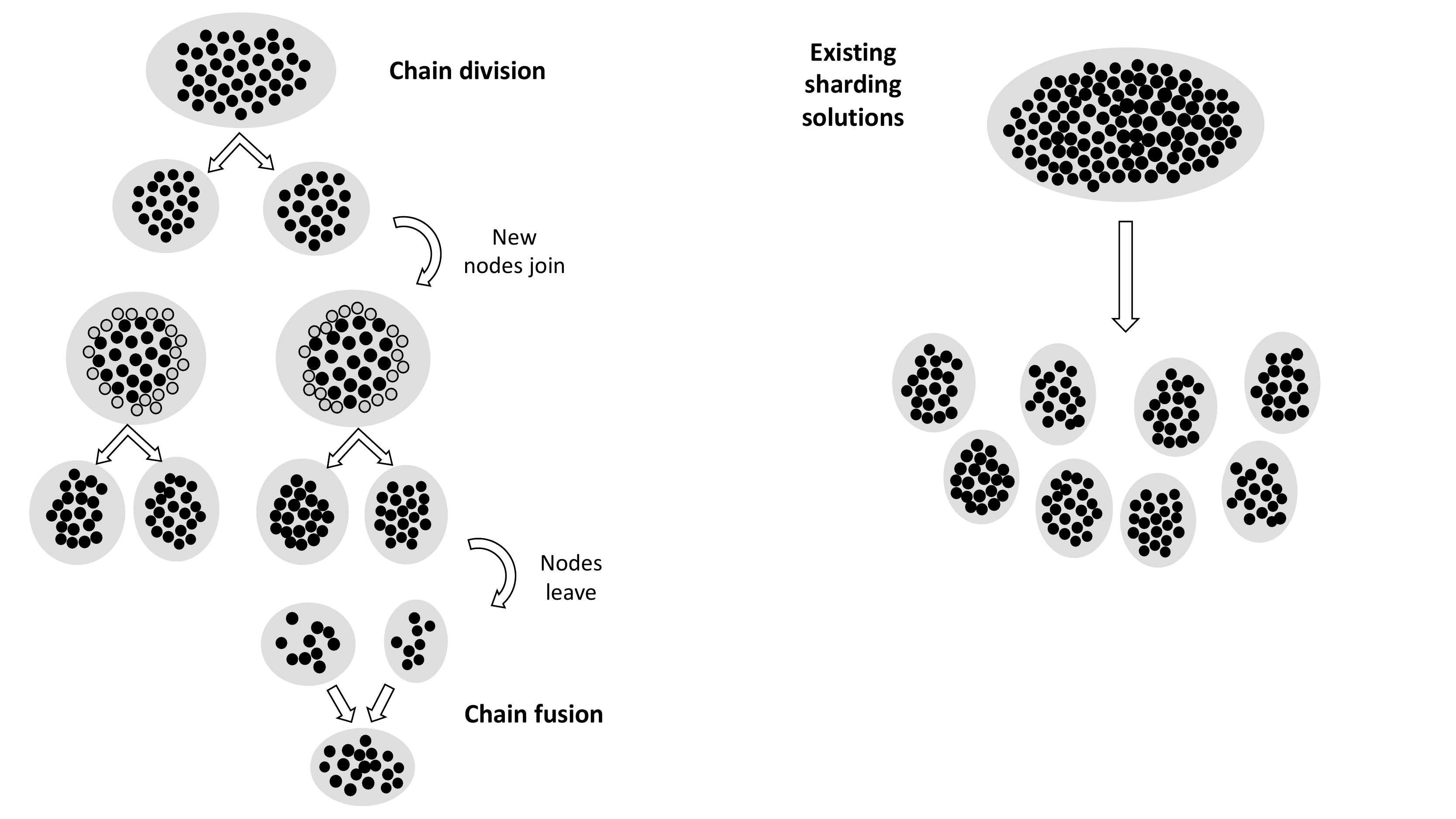}
\caption{Evolution of the validator sets in the case of \our{} (left) and standard sharding solutions (right). Solid bullets indicate existing nodes while and empty bullets denote newly joining nodes.
}
\label{fig:chain:division:vs:sharding}
\end{figure*}

\subsection{Overview of \our}

In \our, we envision a blockchain ecosystem with multiple blockchains running autonomously.
Each chain~$C$ comprises a set of users~$\mathcal{U}_C$ sharing a given business logic.
The clients of a chain~$C$ issue transactions, which are included to the transaction ledger~$\ledger_C$---distributed across blockchain nodes.
The ledger is an ordered sequence of transactions, agreed upon by the validators running a dedicated consensus protocol.
We denote by~$V_C \subseteq \mathcal{U}_C$ the validator set of chain~$C$.
In the rest of the paper, we refer to the size of the validators set, denoted by~$\size{V_C}$,  as the \emph{size of chain}~$C$.
Every validator~$v\in V_C$ is expected to store its own local copy~$\ledger_{C,v}$ of the ledger, and to participate in the consensus protocol for extending the ledger with new transactions.

Each chain in \our{} can adopt its own consensus protocol regardless of the choice of other chains, operating as an autonomous system. Different blockchains can however interact with each other, e.g., to transfer assets across different chains.
Interoperability among the various chains is enabled through dedicated functionalities that let a chain read from, or (conditionally) write to, the state of another chain.
Our system allows blockchains to form and evolve dynamically as new members join the system.

At the core of \our{} is a mechanism that lets the various blockchains
to self-organize and dynamically create new sibling chains, as the need comes.
Essentially, an existing blockchain may trigger the division of itself in order to increase throughput by parallelising the processing of transactions.
In contrast to blockchain sharding, where ``parallel processing'' translates to the various shards splitting the load of transaction processing under the same consensus, our solution can be seen as a way to realize fully autonomous shards which operate independently of each other and possibly under different consensus protocols.
Moreover, our system triggers chain division dynamically and only for those chains experiencing a performance congestion, therefore offering higher flexibility.
In Figure~\ref{fig:chain:division:vs:sharding} we provide a high-level illustration of how (the set of validators of) the blockchains in our system evolve, as new participants join, compared to blockchain sharding.
\our{} enables parallelising the processing of transactions, as in sharding, while also keeping the size of each shard small (in our case, a ``shard'' consists of an autonomous permissioned blockchain).

Here, the main obstacle to keeping the shard size small when splitting is to prevent faulty nodes from concentrating in one shard, as this may lead to a violation of the consensus bounds.
This is the main challenge that all sharding systems need to overcome.
To do so, existing sharding techniques refresh all shards periodically, running dedicated reconfiguration protocols. In contrast, \our{} triggers chain division recursively, creating two sibling chains at a time, and a new division is triggered only once a given chain has become sufficiently large (cf. Figure~\ref{fig:chain:division:vs:sharding}).
Crucially, a chain becomes ``sufficiently large'' by extending its set of nodes, roughly doubling in size, and the newly added nodes have the effect of re-balancing the faulty ratio, thereby ensuring that division does not compromise robustness.
More specifically, the new nodes who join the system are faulty according to a given ratio~$\frac{f}{n}$---strictly below the threshold tolerated by the consensus protocols adopted in the chains.
Therefore, even if one chain-splitting operation led to the creation of a ``more faulty'' chain, i.e., with a faulty ratio slightly above~$\frac{f}{n}$ (which can happen with a small probability), doubling the size of such chain by adding new nodes pushes the ratio ``back'' to~$\frac{f}{n}$.

We proceed with describing the various routines to create and evolve chains in our system in Section~\ref{sec:solution:chain:management}.
In Section~\ref{sec:solution:crosschain:communication}, we also discuss how to enable cross-chain communications among the various chains in the system.

\subsection{Chain Management}
\label{sec:solution:chain:management}

\noindent\textbf{Chain creation.}
This process enables the creation of a blockchain ``from scratch'' (in contrast to creating it via chain division, which we describe later). It requires setup among a set of users to establish the configuration of a new blockchain, which include:
a unique identifier~$C$ for the chain,
a set of validators~$V_C$,
the specification of a consensus protocol~$\Pi_C$,
a set~$\mathcal{C}_C$ of clients (where $\mathcal{U}_C = \mathcal{C}_C \cup V_C$ provides all users in~$S$), and
the initial distribution of assets~$\vec{A}[u]_{u\in\mathcal{U}_S}$ among users in~$C$.
As \our{} is application-agnostic, we abstract away this phase and  declare a chain~$C$ to be created upon request of the relevant (registered) validators in~$V_C$. Compactly,
\begin{equation}
    \ledger_C[0] \gets \mathsf{ChainCreation}(V_C,\langle \mathit{config} \rangle)
\end{equation}
indicates the creation of chain~$C$, with validator set~$V_C$ and configuration specified in~$\langle \mathit{config}\rangle$. Upon completion of this phase, all validators in~$V_C$ are provided with the genesis block~$\ledger_C[0]$---which summarizes pertinent information about the chain.
They can hence start running the blockchain and extending the ledger.

\vspace{1 em}\noindent\textbf{Joining a chain.}
This procedure is necessary when a user~$u$ wishes to become a member of a given chain~$C$. For clients, it is sufficient to submit a registration transaction directly to the validators~$V_C$. However, if~$u$ wants to become a validator of~$C$, the request is first examined by the existing validators against some pre-established access-control policies and, if the request fulfils the policies, a configuration update is triggered for including~$u$ to~$V_C$, leading an actual reconfiguration step to update~$V_C\gets V_C \cup \{u\}$. The update is recorded in the registry.

\vspace{1 em}\noindent\textbf{Chain division.}
This is the core procedure of our system, and the most crucial for security (as we analyse in Section~\ref{sec:analysis}).
It enables the division of a blockchain~$C$, dubbed \emph{parent chain}, for ``giving birth'' to two new chains~$C_1$ and~$C_2$, the \emph{child chains}. Concretely, it triggers the partitioning of the validator set $V_C = V_{C_1} \cup V_{C_2}$, so that the validators in~$V_{C_i}$ become members of child chain~$C_i$.
In this way, it maintains manageable validator sets with optimal size and, therefore, optimal throughput.
Chain division is inspired by the biological process of cell division, i.e., \textit{mitosis}, that creates two child cells from a parent cell.

The chain-division process may be triggered by various events, depending on the application scenario and consensus implementation.
For instance, the system could be set to support a maximum chain size~$n_{max}$:
in this case, chain division is requested as soon as one of the system's chains~$C$ reaches the pre-established maximum size $\size{V_C} \geq n_{max}$.
Alternatively, the system could monitor the transaction throughput of each chain and trigger chain division in case the measurements show a significant and long-lasting throughput drop.
Our system does not make any restriction in this regard.
Since our solution is application-agnostic, we abstract away the specific mechanism used and assume an implicit triggering event.

We specify the instructions for each validator~$v_i\in V$ to execute the chain division process in Algorithm~\ref{algo:chain:division}.
\begin{algorithm}[tb]
    \caption{Chain Division process.}
    \label{algo:chain:division}
    \begin{algorithmic}[1]
        \Procedure{Chain Division}{$C,V$}
        \Comment{Code for~$v_i\in V$}
        \State upon $\mathit{trigger\; event}$ do:%
        \label{algo:line:trigger}
        \State \quad send $\langle \textsc{Divide},C,v_i\rangle$ to all~$v_j\in V$
        \State upon deliver $\langle \textsc{Divide},C,\textrm{initiator}\rangle$ do:
        \State \quad Verify $\mathit{trigger\; event}$
        \State \quad if $\textrm{initiator}\neq v_i$ do:
        \State \qquad Verify $\textrm{initiator}\in V$
        \State \quad $s_i \gets \mathsf{Sign}(\mathrm{pk}_i,\textsc{Divide},C,\textrm{initiator})$
        \State \quad send $(\langle\textsc{Divide},C,\textrm{initiator}\rangle,s_i)$ to all $v_j\in V$%
        \label{algo:line:ack}
        \State \quad $\mathit{acks} \gets \emptyset$
        \State upon deliver $(\langle\textsc{Divide},C,\textrm{initiator}\rangle,s_j)$ from $v_j$ do:
        \State \quad Verify $s_j$ for message $\langle\textsc{Divide},C,\textrm{initiator}\rangle$
        \State \quad $\mathit{acks} \gets \mathit{acks}\cup \{s_j\}$
        \State upon $\size{\mathit{acks}} \geq \mathit{quorum}$ do:%
        \label{algo:line:ack:sufficient}
        \State \quad trigger $(V_1,V_2)\gets\mathsf{ChainDivision.Assign}(V)$
        \State \quad trigger $(C_1,C_2)\gets\mathsf{ChainDivision.Reconfig}(C)$
        \EndProcedure
    \end{algorithmic}
\end{algorithm}

Let~$C$ be the parent chain and let~$V$ be its validator set.
The division process is initiated by any validator~$v_i$ that, upon observing the pre-established condition for division (i.e., $\mathit{trigger\; event}$ in line~\ref{algo:line:trigger}), issues a division request to all validator in~$V_C$.
As soon as a validator~$v_j$ receives the division request, it verifies that the triggering event happened and, if this is the case, broadcast an acknowledgment to proceed with chain division (cf.~line~\ref{algo:line:ack}). The acknowledgment is a signature, under $v_j$'s registered signing key, of the division request.
After sufficiently many acknowledgements have been collected (cf.~line~\ref{algo:line:ack:sufficient}),
specifically, at least~$\mathit{quorum}$ many depending on the consensus protocol (a typical choice could be $\mathit{quorum} \geq (1-\alpha)\size{V}$ where $\alpha$ is the tolerated failure threshold),
chain division proceeds with the actual split of the chain~$C$ and its validator set~$V_C$.

This second phase comprises:
a \emph{validator assignment scheme} denoted by~$\mathsf{ChainDivision.Assign}$), to split~$V$ into two subsets~$V_1$ and~$V_2$, the validator sets for the child chains to be created;
and a \emph{reconfiguration protocol} denoted by~$\mathsf{ChainDivision.Reconfig}$, for replacing the original blockchain~$C$ with two new chains~$C_1$ and~$C_2$ and making sure the child chains are initialized consistently with the parent chain.
Compactly:
\begin{align}
    (V_{1},V_{2}) &\gets \mathsf{ChainDivision.Assign}(V)\\
    (C_1,C_2) &\gets \mathsf{ChainDivision.Reconfig}(C)
\end{align}
Algorithm $\mathsf{ChainDivision.Assign}$ defines a method to partition~$V$ into~$V_{1}$ and~$V_{2}$.
For robustness purposes, the assignment method should be robust in the sense of ensuring that for each child chain, the faulty participants in~$V_i$ are below~$\alpha_i \size{V_i}$, where~$\alpha_i$ is the tolerated failure threshold for the consensus protocol of chain~$C_i$, i.e., the maximum tolerated fraction of non-correct participants.
The assignment of validators is crucial for robustness in the presence of faulty processes, as we discuss in greater detail in Section~\ref{sec:division:validator:assignment} (and analyse formally in Section~\ref{sec:analysis}).

Once the assignment of validators to sets~$V_1$ and~$V_2$ has been established, the validators in~$V_{i}$ set up a new blockchain~$C_i$; this step is similar to the creation of a new chain ``from scratch'', with the exception that both blockchains~$C_1$ and~$C_2$ must be consistent with the state of their parent blockchain~$C$.
Namely, the state of each child chain~$C_i$ is fully described by its ledger~$\ledger_{C_i}$, and the latter reproduces all information about validators and assets registered in~$\ledger_C$, for all validators~$v\in V_{C_i}$.
Notice that, to ensure consistency of the child ledgers with the parent ledger, it is necessary that all (honest) validators agree on the state of ledger~$\ledger_C$ prior to initiating the reconfiguration protocol.

Upon initialization of both child chains~$C_1$ and~$C_2$, the corresponding validators issue a configuration-update request to the membership service, so that the division of chain~$C$, with corresponding creation of chains~$C_1$ and~$C_2$, is registered.
Upon completion of this step, the newly created chains can start operating.
From this moment on, they proceed autonomously and independently of each other.

\vspace{1 em}\noindent\textbf{Chain fusion.}
Complementing the chain-division procedure, \our{} also supports a fusion operation that creates a single set of validators~$V'$ by merging two existing validator sets~$V_1$ and~$V_2$. Similarly to chain division, this operation triggers a reconfiguration step aimed at replacing the two blockchains~$C_1$ and~$C_2$ with a new blockchain~$C'$.
Unlike chain division, chain fusion does not present any particular challenge in terms of robustness. However, when combining heterogeneous chains that use different consensus protocols, in particular, that guarantee correctness for different failure thresholds~$\alpha_1 \neq \alpha_2$, the resulting merged chain will be resilient to the smallest failure ratio, i.e., ~$\alpha' \leq \min \{\alpha_1,\alpha_2\}$.

\subsection{Validator Assignment Scheme}
\label{sec:division:validator:assignment}

The validator assignment scheme is a crucial subroutine of the chain division process (cf.~Section~\ref{sec:solution:chain:management}), as it determines which validators in the parent chain~$C$ are assigned to which of the child chains.
Its design depends on the ratio~$\beta$ of (crash- or Byzantine) failures in the parent chain (i.e., $f = \beta \size{V} < \alpha \size{V}$), as well as on the tolerated thresholds~$\alpha_1$ and~$\alpha_2$ for the two child chains. Namely, the assignment scheme must ensure the ratio of failing participants is below~$\alpha_i$ in both child chains.
We discuss two alternatives, a deterministic assignment and a randomized assignment.

\vspace{1 em}\noindent\textbf{Deterministic Assignment.}
The robustness condition is automatically fulfilled by assuming a more conservative bound on the fraction of tolerated faults in~$V$. Namely, if the consensus protocol run by chain~$C_i$ can tolerate a number of faulty nodes below~$\alpha_i \size{V_i}$, then requiring a bound~$f < \frac{\alpha_i}{2} \size{V}$ in the parent chain, where $f$ is the number of faulty nodes in~$V$, suffices to guarantee robustness in both child chains.
That is, the conditions for consistency and liveness are met regardless of how the validators are assigned to the child chains.
One such deterministic assignment scheme could simply rank validators in~$V$ (e.g., following the lexicographic order of the validator's identifiers), hence assign the first~$n/2$ validators in the ranking to~$V_1$, and the other validators to~$V_2$.

\vspace{1 em}\noindent\textbf{Randomized Assignment.}
An alternative method to assign validators to the child chaims could leverage randomization to distribute the failing nodes between~$V_1$ and~$V_2$ according to the ratio~$\alpha_1 : \alpha_2$, so that the failure ratios are preserved with high probability.
Compared to a deterministic assignment scheme, using randomization allows to tolerate a higher number faults in~$C$, at the price of providing probabilistic security guarantees. We analyse sufficient conditions for the security of chain division under a randomized assignment scheme in Section~\ref{sec:analysis}.
Crucially, in the case of malicious nodes controlled by an attacker, the scheme must also prevent Byzantine nodes from biasing the randomness used---to prevent the attacker from influencing the selection of validators and gather all of its nodes in one chain (we call this a ``Byzantine gathering'').
One such robust assignment scheme could rank validators based on publicly available randomness extracted from the blockchain, e.g., by cryptographically hashing each validators' identifier with recent blockchain content.

\subsection{Cross-chain Communication}
\label{sec:solution:crosschain:communication}

So far, we discussed the relevant routines to manage the various chains in the system, at the blockchain level, which only involve the validators maintaining the platform.
In the sequel, we discuss the services provided by~\our{} to \emph{clients}.
Besides standard transaction processing within one chain, our system support the communication between different chains, thereby letting clients issue cross-chain transactions.
Concretely, we provide two smart-contract functionalities that clients can use to communicate with (the clients of) other chains: \emph{transfer of knowledge} and \emph{transfer of asset}.

\vspace{1 em}\noindent\textbf{Transfer of Knowledge (ToK).}
This functionality allows proving a statement, defined through a predicate~$P$, about a given blockchain.
The predicate can be evaluated on a source chain~$C_s$ using local information, and is then shown to be correct on a target chain~$C_t$.
The ToK protocol is run between two participants, a client~$\prover$ that acts as a prover, and a third party (another client or a blockchain) in the role of a verifier~$\verifier$.
It is defined by the following two algorithms:
\begin{enumerate}
    \item $\ToKGenerateProof(P,\ToKtag)$, which receives as input a predicate $P$ and a tag~$\ToKtag$, and returns a valid proof~$\ToKproof$ if the predicate is true, otherwise it returns an error~$\bot$;
    \item $\ToKVerifyProof(\ToKproof,\ToKtag)$, which receives as input a proof~$\ToKproof$, a tag~$\ToKtag$, and returns a verdict~$v\in\{0,1\}$ about the validity of the proof.
\end{enumerate}
The protocol is as follows: $\prover$ retrieves a freshness tag~$\ToKtag$ from~$\verifier$ and invokes~$\ToKGenerateProof(P, \ToKtag)$ for a pre-established predicate~$P$. If this invocation returns an error, the transfer of knowledge fails---either the predicate is incorrect, or it requires information which is unavailable to~$\prover$.
Otherwise, a proof~$\ToKproof$ for the validity of~$P$ is generated, hence~$\prover$ can forward~$\ToKproof$ to $\verifier$.
The verifier finally invokes~$\ToKVerifyProof(\ToKproof,\ToKtag)$
to validate the provided proof. A negative outcome means that either the predicate is not valid (specifically: no valid quorum in chain~$C_s$ has signed the predicate) or it is not fresh (i.e., the tag is no longer valid).

\vspace{1 em}\noindent\textbf{Transfer of Asset (ToA).}
This functionality enables the transfer of a given asset from one chain to another.
We build ToA based on Transfer of Knowledge, using an additional locking mechanism.
We define ToA through the following algorithms:
\begin{enumerate}
    \item $\ToALock(a,A_t,C_t)$, which locks an asset~$a$ (on source chain $C_s$) so that it can be later retrieved on target chain~$C_t$ by address~$A_t$. Invoking this function  makes the asset temporarily unavailable in chain~$C_s$, until the transfer is resolved. A successful call to~$\ToALock$ triggers a transfer of knowledge about the inclusion of a lock transaction in chain~$C_s$, and it generates a proof~$\ToAprooflock$ that the asset has been locked.
    \item $\ToAClaim(\ToAprooflock)$, which verifies the proof~$\ToAprooflock$ and, if the proof is valid, it creates the associated asset in chain~$C_t$ and links it to address~$A_t$.
    Regardless of whether the proof is valid or not (i.e., the verdict is either~$v=1$ or~$v=0$), the corresponding transaction along with~$v$ is recorded into chain~$C_t$. A ToK returns a proof~$\ToAproofclaim$ in case of success, or a proof~$\ToAproofabort$ if the claim has failed.
    \item $\ToAResolve(\pi)$, which completes the ToA process by either rolling back the locking in case of abort ($\pi = \ToAproofabort$), or deleting the locked asset in~$C_s$ (if $\pi = \ToAproofclaim$). Invalid proofs are discarded, and do not change the state of the locked asset.
\end{enumerate}

We will describe how to integrate the two functionalities just described within existing permissioned blockchains in Section~\ref{sec:fabric}.
\section{Security Analysis}
\label{sec:analysis}

In this section, we analyse the security of \our{} for the case of a randomized assignment of validators, identifying sufficient conditions for preventing gatherings of faulty nodes (cf.~Section~\ref{sec:division:validator:assignment}).
Specifically, we determine how many faulty participants can be tolerated in the parent chain to ensure that violations of the security bounds in the child chains are unlikely.

Let~$V$ be the validator set of the parent chain, let~$\size{V} = n$, and let~$V_1$ and~$V_2$ denote the validator sets of the derived child chains, with~$n_i = \size{V_i}$.

Let~$f$ denote the number of faulty nodes in~$V$.
We assume that the number of faulty participants in the parent chain is bounded by a constant fraction of the overall participant, i.e., $f < \alpha n$ for some protocol-specific threshold~$0\leq \alpha\leq \frac{1}{2}$.
For instance, $\alpha = \frac{1}{3}$ for asynchronous BFT consensus, meaning that the protocol can tolerate up to~$33\%$ Byzantine faults.%
\footnote{Different protocols might tolerate a different fraction of faulty nodes, for instance, $\alpha = 1/2$ suffices for synchronous BFT protocols and asynchronous CFT protocols.}
The security of chain division requires the consensus bound to hold for the child chains, too: if~$f_i$ denotes the number of faulty nodes in~$V_i$, then we require $f_i < \alpha_i n_i$ for $i=1,2$, where~$\alpha_1$ and~$\alpha_2$ depend on the consensus protocols run by the individual chains.

In the rest of this section, we assume for simplicity the same consensus bounds for all three chains, i.e., $\alpha = \alpha_1 = \alpha_2$, and let chain division split~$V$ into two halves, i.e., $n_i = \frac{n}{2}$.
Notice that if the number of faults in~$V$ is just below the tolerated threshold, i.e., $f = \lfloor \alpha (n-1)\rfloor$, preserving security after splitting requires that the sibling chains contain exactly~$f_i = \frac{f}{2}$ faulty nodes each.
Although a randomized assignment is likely to distribute the faulty nodes equally among the two sibling chains, statistical fluctuations may create an imbalance---leading to violating the consensus bound in one of the sibling chains.
In contrast, if the fraction of faulty participants in~$V$ is sufficiently small---strictly below the consensus bound---the sibling chains still meet the consensus bounds with high probability.%
\footnote{A similar argument applies to the committee selection in sharding protocols.}
Below we formalize this intuition.

\begin{figure*}[th]
    \centering
    \begin{subfigure}[b]{0.49\textwidth}
    \centering
    \begin{tikzpicture}
        \begin{axis}[
            forget plot style={opacity=1},
            xlabel={$\beta$},
            ylabel={Prob.},
            domain=0.0:0.33,
            grid=major,
            legend entries={
                $n=10$,
                $n=40$,
                $n=50$,
                $n=100$
                },
            legend pos=north west,
            ]
            \addplot+[mark=,thick,blue,dashed] [forget plot]{exp(-10*(1/3 - x)^2)};
            \addplot[very thick, mark=*,blue] table [x = $x$, y = $y$]{hypergeom_0.33_10.dat};
            \addplot+[mark=,very thick, red,dashed] [forget plot]{exp(-40*(1/3 - x)^2)};
            \addplot[very thick, mark=square,red] table [x = $x$, y = $y$]{hypergeom_0.33_40.dat};
            \addplot+[mark=,very thick, brown,dashed] [forget plot]{exp(-50*(1/3 - x)^2)};
            \addplot[very thick, mark=*,brown] table [x = $x$, y = $y$]{hypergeom_0.33_50.dat};
            \addplot+[mark=,very thick, black,dashed] [forget plot]{exp(-100*(1/3 - x)^2)};
            \addplot[very thick, mark=star,black] table [x = $x$, y = $y$]{hypergeom_0.33_100.dat};
        \end{axis}
    \end{tikzpicture}
    \caption{$\alpha = \frac{1}{3}$.}
    \label{fig:plots:probability:sec:violation:33}
 \end{subfigure}
 \hfill
 \begin{subfigure}[b]{0.49\textwidth}
    \centering
    \begin{tikzpicture}
        \begin{axis}[
            forget plot style={opacity=1},
            xlabel={$\beta$},
            ylabel={Prob.},
            domain=0.0:0.5,
            grid=major,
            legend entries={
                $n=10$,
                $n=40$,
                $n=50$,
                $n=100$
                },
            legend pos=north west,
            ]
            \addplot+[mark=,very thick, blue,dashed]  [forget plot]{exp(-10*(1/2 - x)^2)};
            \addplot[very thick, mark=*,blue] table [x = $x$, y = $y$]{hypergeom_0.5_10.dat};
            \addplot+[mark=,very thick, red,dashed]  [forget plot]{exp(-40*(1/2 - x)^2)};
            \addplot[very thick, mark=square,red] table [x = $x$, y = $y$]{hypergeom_0.5_40.dat};
            \addplot+[mark=,very thick, brown,dashed]  [forget plot]{exp(-50*(1/2 - x)^2)};
            \addplot[very thick, mark=*,brown] table [x = $x$, y = $y$]{hypergeom_0.5_50.dat};
            \addplot+[mark=,very thick, black,dashed]  [forget plot]{exp(-100*(1/2 - x)^2)};
            \addplot[very thick, mark=star,black] table [x = $x$, y = $y$]{hypergeom_0.5_100.dat};
        \end{axis}
    \end{tikzpicture}
    \caption{$\alpha = \frac{1}{2}$.}
    \label{fig:plots:probability:sec:violation:50}
\end{subfigure}
    \caption{Probability of violating security when~$\alpha_i = \frac{1}{3}$ (Figure~\ref{fig:plots:probability:sec:violation:33}) and $\alpha_i =\frac{1}{2}$ (Figure~\ref{fig:plots:probability:sec:violation:50}), and $\beta < \alpha_i$. For completeness, we also plot the upper-bound limit for each probability in dotted lines).}
    \label{fig:plots:probability:sec:violation}
    \end{figure*}
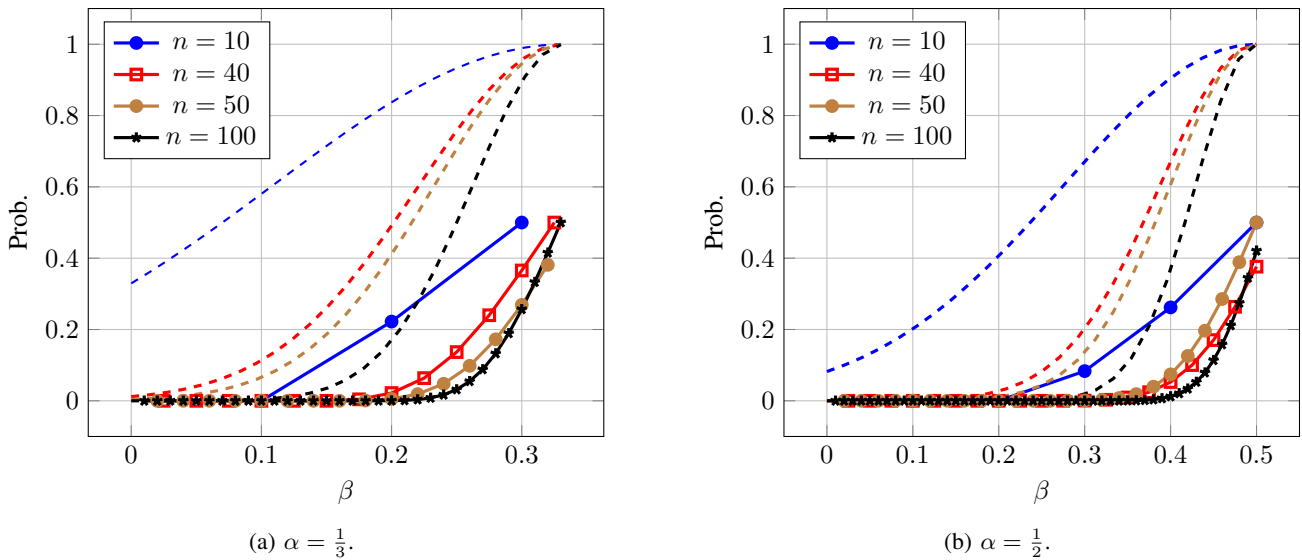

Let~$f = \beta n$ be the actual number of faulty participants in the parent chain, with~$\beta < \alpha$.
We assume a randomized process that assigns each validator in~$V$ to either of the sets~$V_1$ and~$V_2$ uniformly at random, subject to the restriction that~$n_1 = n_2 = \frac{n}{2}$.
For the sake of analysis, we consider~$\frac{n}{2}$ nodes being selected uniformly at random from~$V$ (without replacement) and assigned to~$V_1$, and have the remaining nodes assigned to~$V_2$.
Then the number of faulty nodes assigned to each sibling chain, $f_1$ and~$f_2$, are dependent random variables with relation~$f_1 + f_2 = f$ and hypergeometric distribution $f_i \sim H(n,f,\frac{n}{2})$, which we recall below. 

Table~\ref{table:analysis:symbols} summarizes the relevant variables and symbols used throughout the analysis.
\begin{table}[tb]
    \centering
    \caption{Relevant variables and symbols.}
    \label{table:analysis:symbols}
    \begin{tabular}{||c c c||}
     \hline
     Variable & Description & Requirements \\ [0.5ex]
     \hline\hline
     $n$ & Number of nodes in $V$ & -- \\
     $\alpha$ & Threshold of tolerated faults in~$V$ & $f < \alpha n$\\
     $f$ & Number of faulty nodes in $V$ & -- \\
     $\beta$ & Actual fraction of faulty nodes in $V$ & $f = \beta n$\\
     $n_i$ & Number of nodes in $V_i$ & $n_1 = n_2 = \frac{n}{2}$\\
     $\alpha_i$ & Threshold of tolerated faults in~$V_i$ & $f_i < \alpha_i \frac{n}{2}$\\
     $f_i$ & Number of faulty nodes in $V_i$ & $f_1 + f_2 = f$ \\
     [1ex]
     \hline
    \end{tabular}
\end{table}

\paragraph*{Hypergeometric distribution}

A hypergeometric experiments can be described through the following variables:
$N$, the overall number of elements;
$M$, the number of elements ``that count'' ($0\leq M \leq N$);
$n$, the number of elements which are extracted, \emph{without repetition}, from the set ($0\leq n \leq N$);
and $H(N,M,n)$, the number of elements that count among the~$n$ which have been extracted~\cite{skala2013hypergeometric}.
Let~$X$ be a random variable with $X\sim H(N,M,n)$.
Then, we have:
\begin{equation}
    \prob{X = k} = \frac{\binom{M}{k} \binom{N-M}{n-k}}{\binom{N}{n}},
\end{equation}
with support
\begin{equation}
    \mathrm{Supp}(X) = \{ \max(0,n+M-N),\dots,\min(n,M)\},
\end{equation}
and with expected value
\begin{equation}
    \ExpVal{}{X} = n\frac{M}{N}.
\end{equation}
For $0 \leq t \leq n\frac{M}{N}$, the following tail bounds hold:
\begin{align}
    \label{hypergeometric:tail1}
    \prob{X \geq \ExpVal{}{X} + t n} &\leq e^{-2 t^2 n}\\
    \label{hypergeometric:tail2}
    \prob{X \leq \ExpVal{}{X} - t n} &\leq e^{-2 t^2 n}
\end{align}

\paragraph*{Probability of violating security}
We are interested in the probability that security is violated in either of the child chains:
\begin{equation}
    \prob{f_1 \geq \alpha n_1 \lor f_2 \geq \alpha n_2}.
\end{equation}
By the relation $f_1 + f_2 = f$, and using the parametrization~$f = \beta n$, we obtain
\begin{equation}
    \prob{f_1 \geq \alpha \frac{n}{2} \lor f_1 \leq \beta n - \alpha \frac{n}{2}}.
\end{equation}
By the assumptions made, we have $\beta n - \alpha \frac{n}{2} < \alpha \frac{n}{2}$, hence the two events $\{ f_1 \geq \alpha \frac{n}{2}\}$ and $\{ f_1 \leq \beta n - \alpha\frac{n}{2}\}$ are disjoint.
Therefore:
\begin{multline}
    \label{eq:probsum}
    \prob{f_1 \geq \alpha \frac{n}{2} \lor f_1 \leq \beta n - \alpha \frac{n}{2}} =\\
    \prob{f_1 \geq \alpha \frac{n}{2}} + \prob{f_1 \leq \beta n - \alpha \frac{n}{2}}.
\end{multline}
We proceed with evaluating the two terms in Equation~\eqref{eq:probsum} separately, observing that~$\ExpVal{}{f_1} = \beta\frac{n}{2}$ and using the tail bounds for the hypergeometric distribution.
For the upper tail, we have:
\begin{align}
    \label{eq:failureprob:bound1}
       \prob{f_1 \geq \alpha \frac{n}{2}}
    &= \prob{f_1 \geq (\alpha - \beta) \frac{n}{2} + \beta \frac{n}{2}}\\
    &= \prob{f_1 \geq \ExpVal{}{f_1} + (\alpha - \beta) \frac{n}{2}}\\
    & \leq e^{-(\alpha - \beta)^2 n}.
\end{align}
We obtain a similar expression for the lower tail:
\begin{align}
    \prob{f_1 \leq \beta n - \alpha \frac{n}{2}}
    &= \prob{f_1 \leq \ExpVal{}{f_1} - (\alpha - \beta)\frac{n}{2}}\\
    &\leq e^{- (\alpha - \beta)^2 n}
    \label{eq:failureprob:bound2}
\end{align}

We observe that for realistic values of~$n$ (i.e., $n \leq 200$), the tail bounds from equations~\eqref{eq:failureprob:bound1}-\eqref{eq:failureprob:bound2} are not tight, leading to a conservative estimation for the probability of violating security.
In other words, the number of tolerated faulty participants in practice is higher than that derived with the analytic bound. This is well visible in Figure~\ref{fig:plots:probability:sec:violation}, where we plot the probability of violating security in a sibling chain, for $n\in \{10, 50, 100, 200\}$ and $\alpha_i \in \{\frac{1}{3},\frac{1}{2}\}$, along with the corresponding upper bounds.
We approximated the exact values using the cumulative probability:
\begin{equation}
    \prob{f_i \geq \alpha_i n} = \sum_{k = \alpha_i \frac{n}{2}}^{\beta n} \frac{\binom{\beta n}{k} \binom{n - \beta n}{\frac{n}{2}-k}}{\binom{n}{\frac{n}{2}}} .
\end{equation}

Based on the analysis above, we observe that chain division is guaranteed to preserve security, with high probability, as long as the actual faulty ratio~$\beta$ is below~$25\%$ when the maximum faulty ratio~$\alpha_i$ is~$33\%$, and similarly, for~$\beta$ below~$40\%$ when~$\alpha_i$ is~$50\%$. Moreover, as the size of the parent chain increases, \our{} can tolerate a higher faulty ratio.

\begin{figure*}[th!]
   \centering
   %
   \begin{subfigure}[t]{0.4\textwidth}
      \centering
      \includegraphics[width=\textwidth]{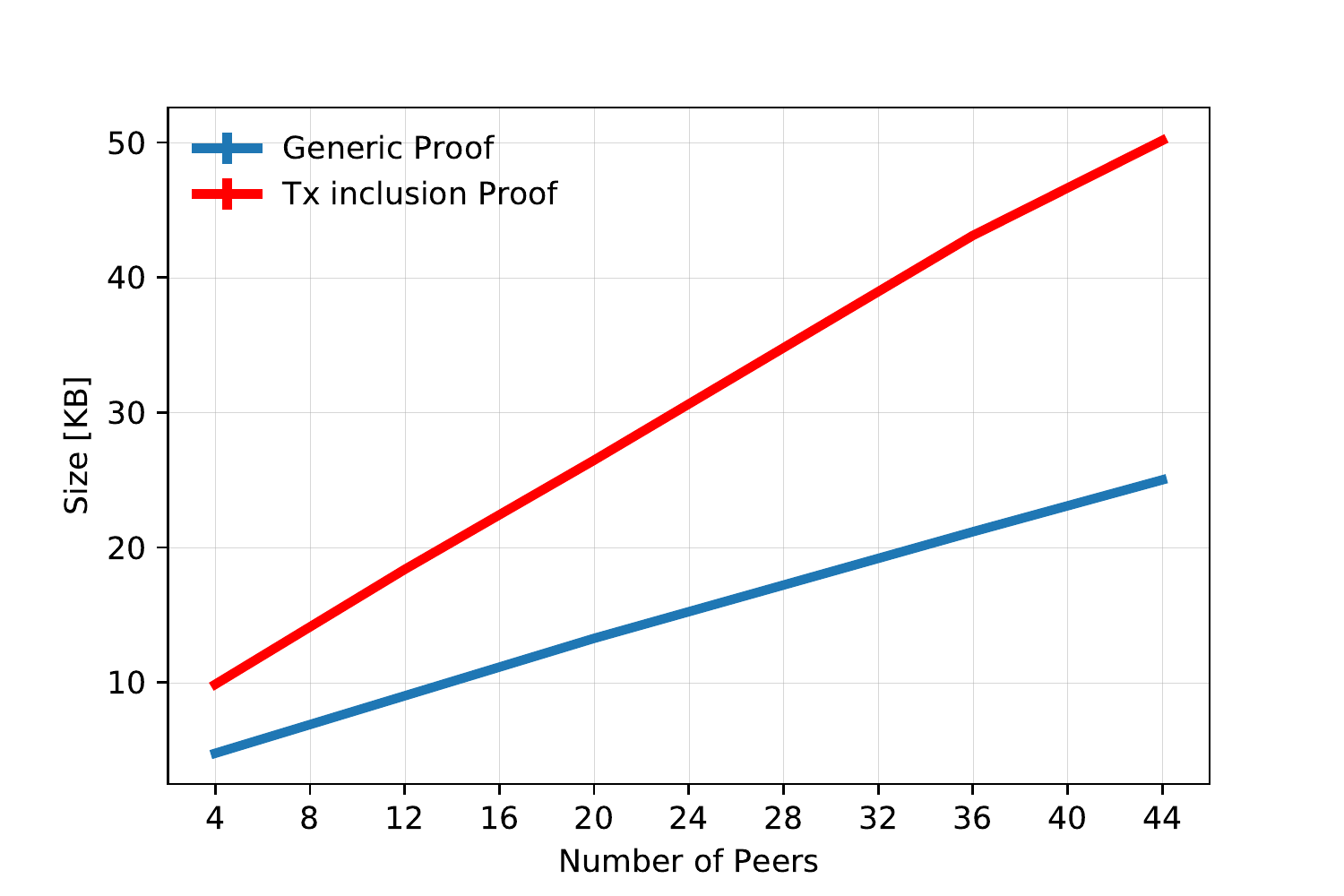}
      \caption{Proof size.}
      \label{fig:proofsize}
       \vspace{1em}
   \end{subfigure}
   \begin{subfigure}[t]{0.4\textwidth}
     \centering
     \includegraphics[width=\textwidth]{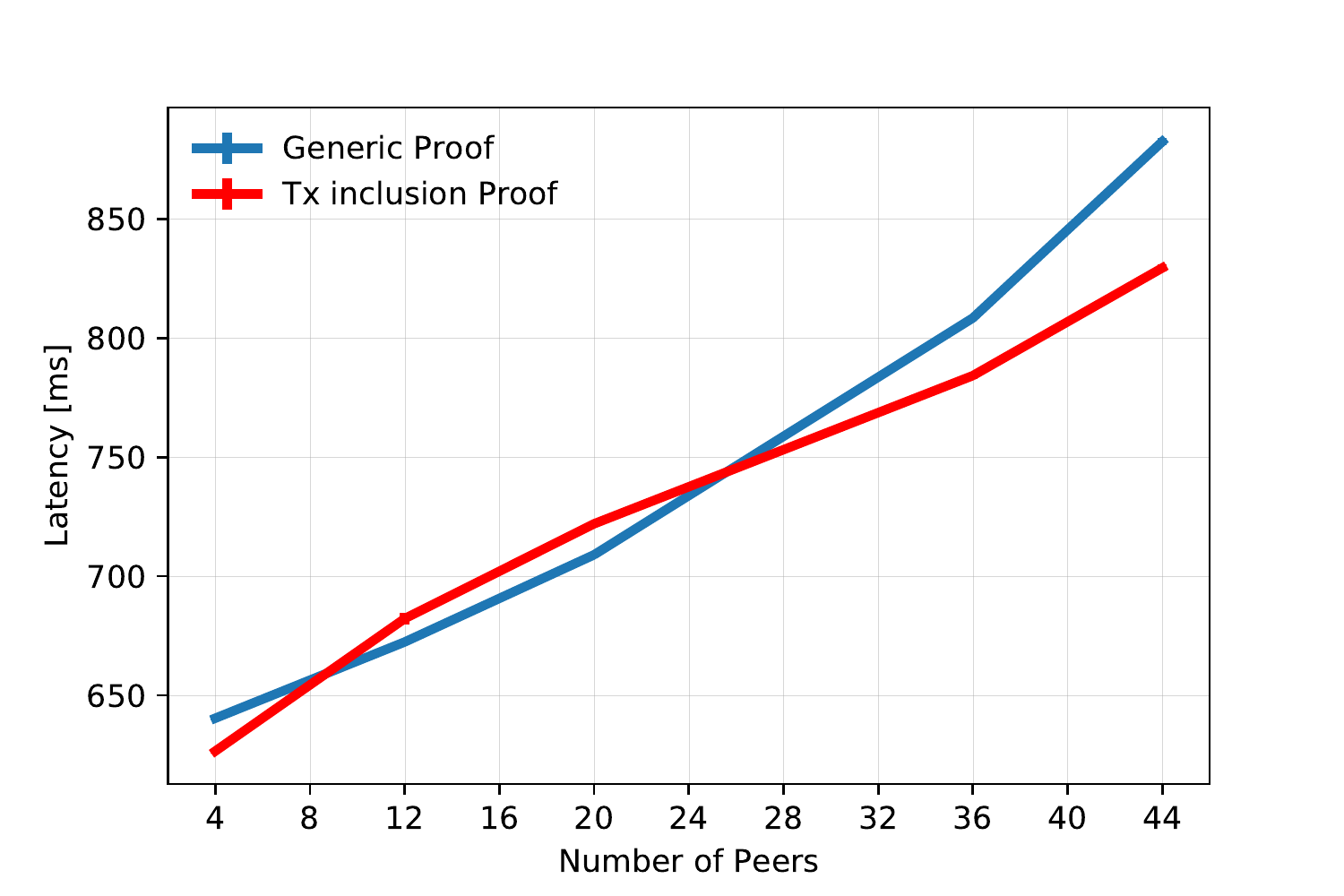}
     \caption{Proof generation latency.}
     \label{fig:proofgentime}
       \vspace{1em}
   \end{subfigure}
   \begin{subfigure}[t]{0.4\textwidth}
      \centering
      \includegraphics[width=\textwidth]{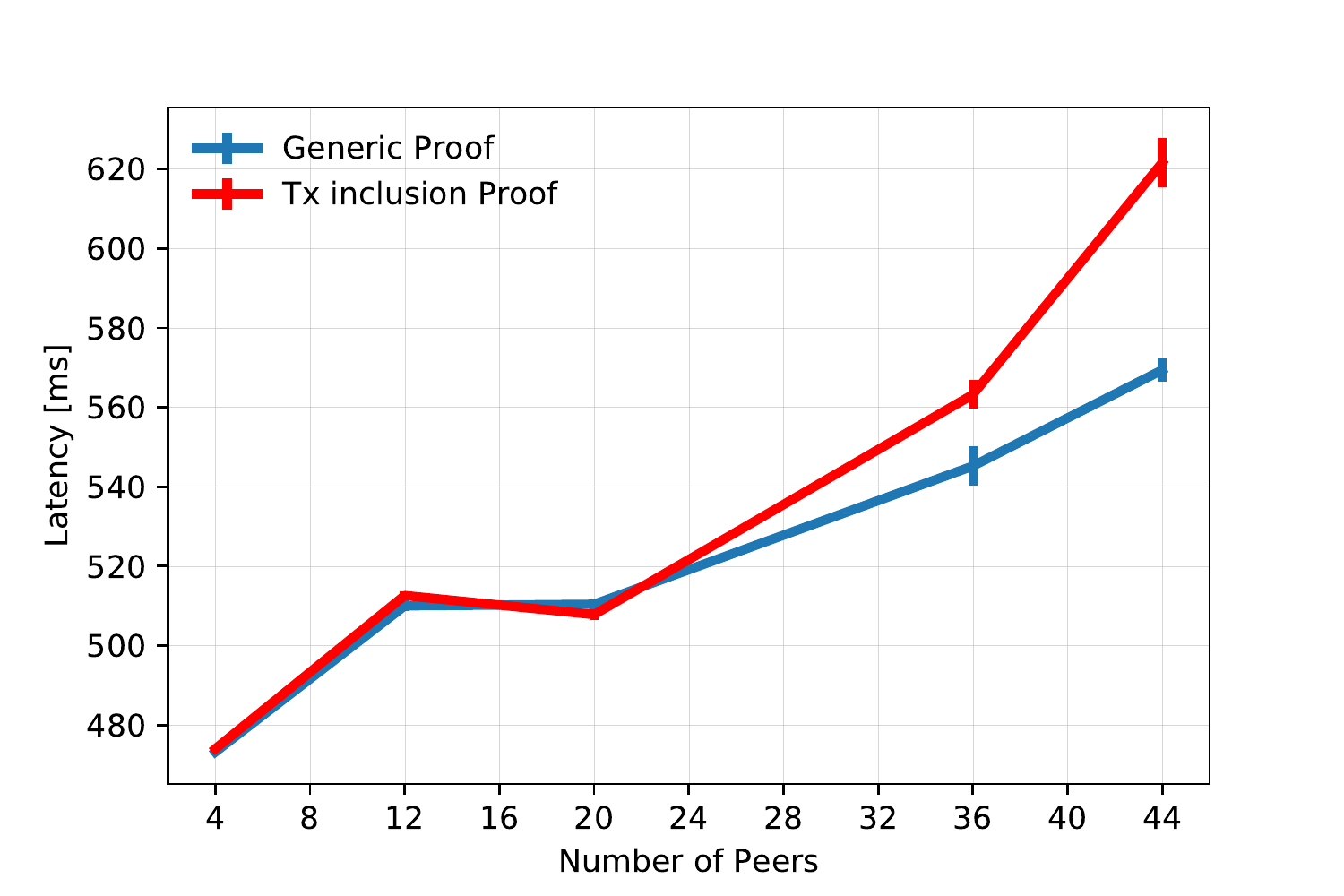}
      \caption{Proof verification latency (off chain).}
      \label{fig:proofverifoff}
        \vspace{1em}
     \end{subfigure}
     \begin{subfigure}[t]{0.4\textwidth}
        \centering
        \includegraphics[width=\textwidth]{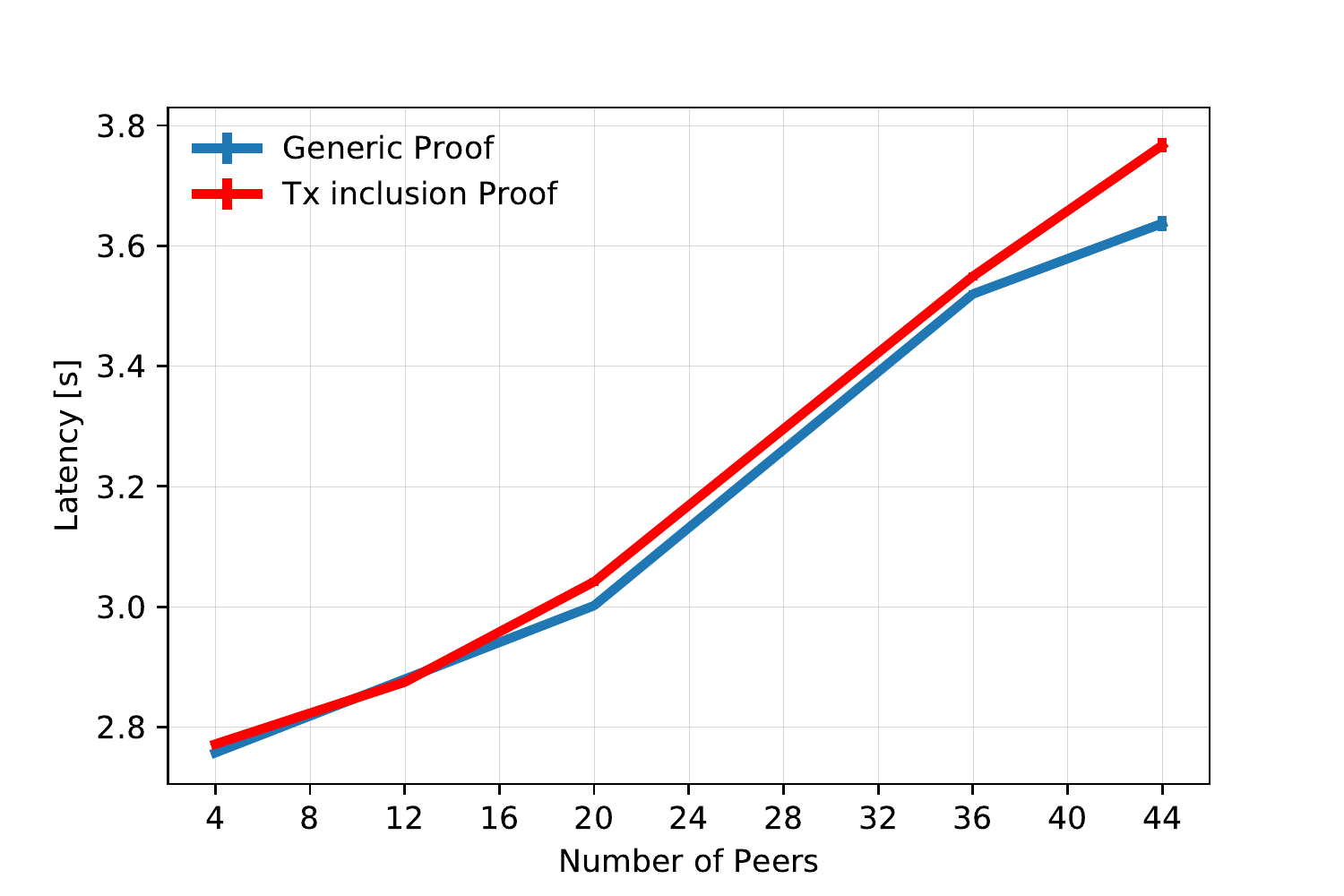}
        \caption{Proof verification latency (on chain).}
        \label{fig:proofverifon}
         \vspace{1em}
     \end{subfigure}
   %
   \begin{subfigure}[t]{0.4\textwidth}
      \centering
      \includegraphics[width=\textwidth]{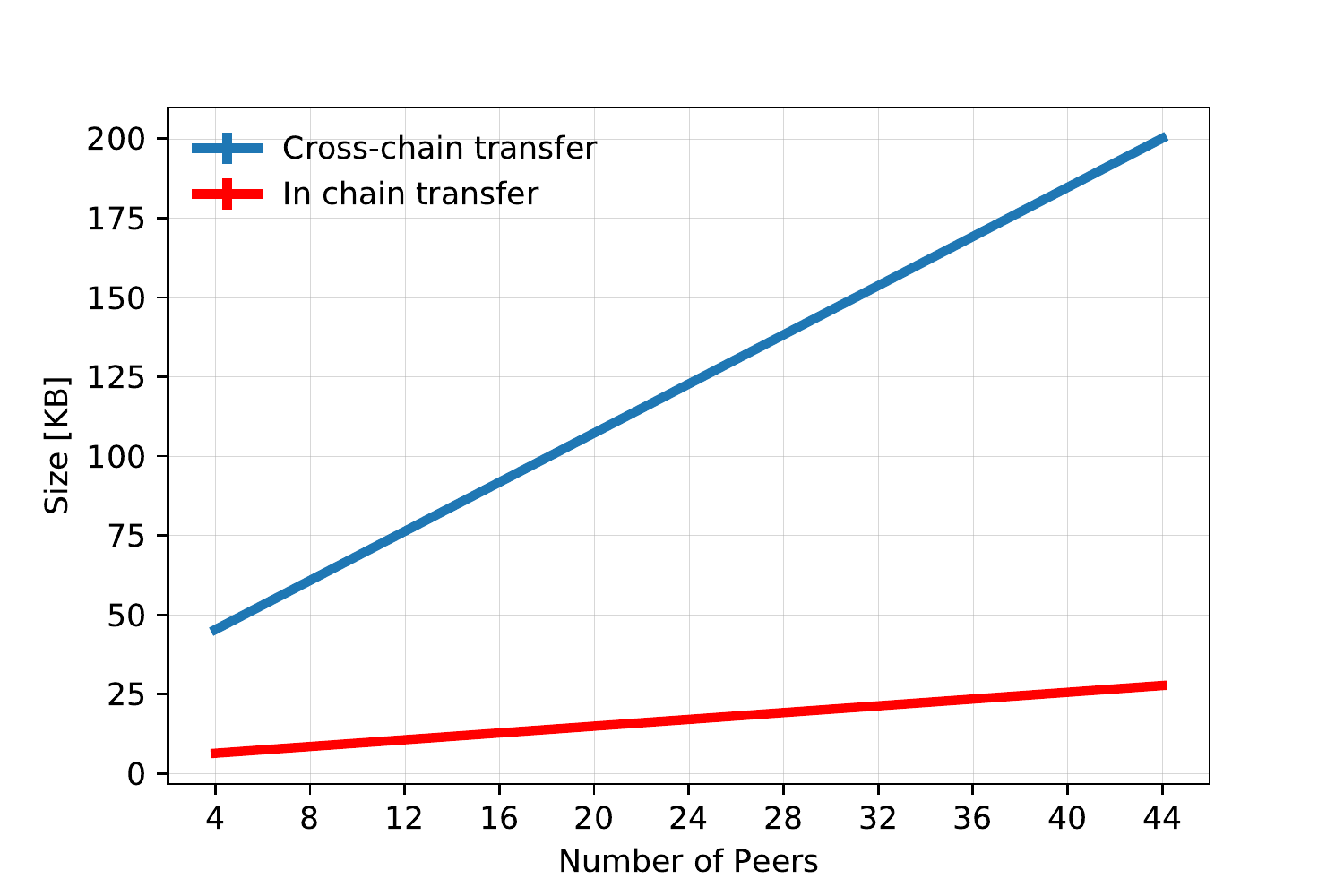}
      \caption{Cross-chain transfer size.}
      \label{fig:ccsize}
   \end{subfigure}
   \begin{subfigure}[t]{0.4\textwidth}
      \centering
      \includegraphics[width=\textwidth]{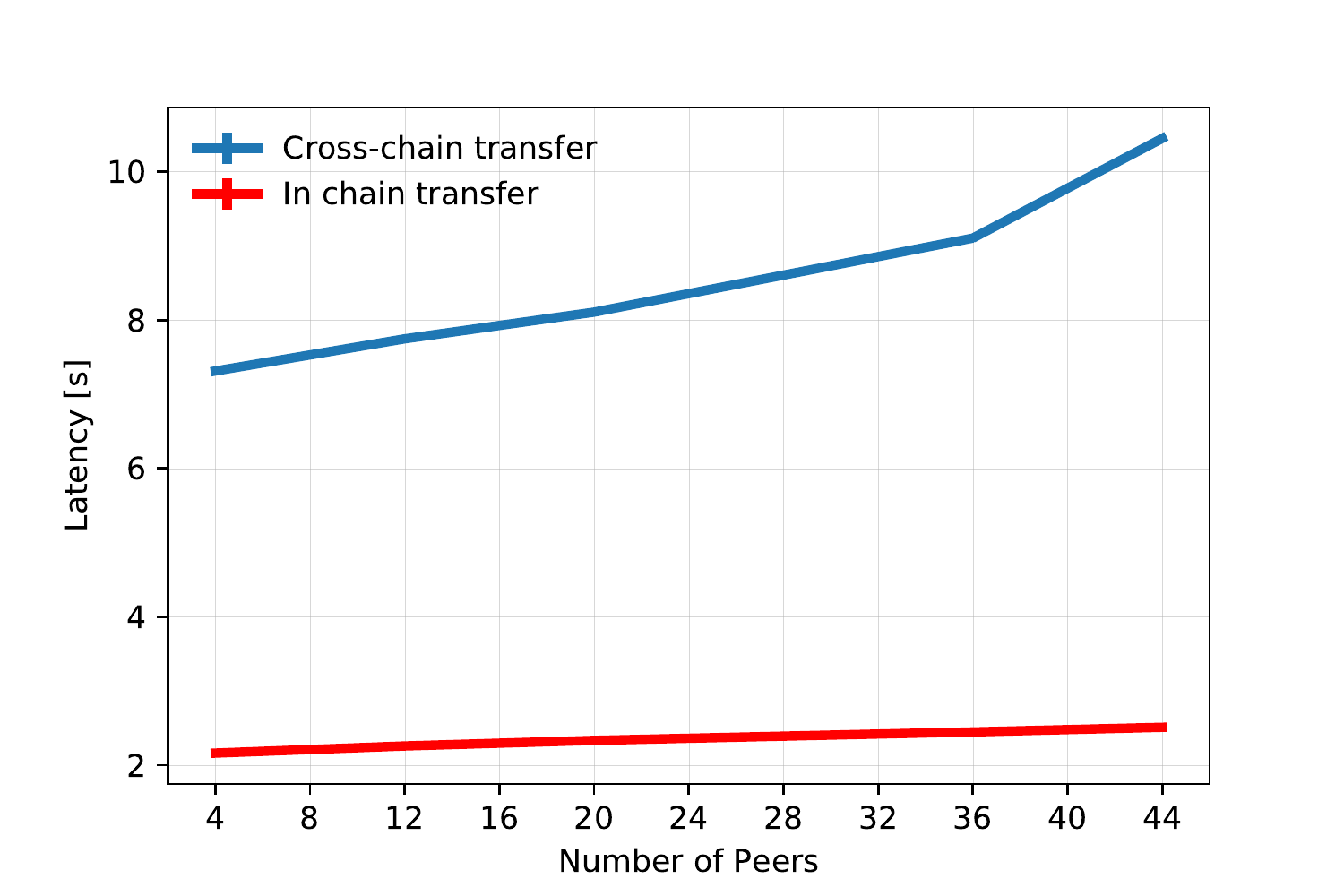}
      \caption{Cross-chain transfer latency.}
      \label{fig:cclatency}
   \end{subfigure}
 \caption{Performance of~\our{} functionalities: size and latency for proof of knowledge generation (Figures~\ref{fig:proofsize}--\ref{fig:proofgentime}), verification (Figures~\ref{fig:proofverifon}--\ref{fig:proofverifoff}), and for cross-chain asset transfer (Figures~\ref{fig:ccsize}--\ref{fig:cclatency}), with respect to an increasing number of peers in each chain.}
 \label{fig:evaluation}
 \end{figure*}

\section{Integration in Hyperledger Fabric}
\label{sec:fabric}

Hyperledger Fabric is a popular and modular operating system for the deployment of permissioned blockchains, developed within an open-source community effort hosted by the Linux Foundation.
Fabric introduces a novel architecture that separates transaction execution from consensus (i.e., transaction ordering). Namely, the Fabric architecture uses the execute-order-validate paradigm, which is in sharp contrast with the traditional order-execute approach used in prior blockchain and state-machine replication (SMR) deployments. Such a paradigm shift is the key enabler for the modularity and flexibility of Fabric. The flexible and modular design of Fabric supports ``pluggable’’ consensus, making it particularly attractive for different applications and use cases.
The Fabric architecture allows expressing flexible trust assumptions: all clients are untrusted (i.e., they are considered as potentially malicious), while peers are grouped into organizations such that mutual trust is assumed within each organization. This model is suitable for accommodating diverse application-specific requirements, as it is the case for our satellite chain ecosystem.
We implemented \our{} on top of Fabric, using Raft~\cite{DBLP:conf/usenix/OngaroO14} for the underlying consensus protocol as it is one of the most widely-deployed CFT protocols. We implemented three functionalities: Transfer of Knowledge, Transfer of Asset, and Chain Division (cf.~Section~\ref{sec:solution}), which we describe below.

We implemented two versions of the Transfer of Knowledge functionality: the first is a generic proof based on predicates evaluated using data known by the chaincode; the second one is a transaction inclusion proof. 

\paragraph*{$\ToKGenerateProof(P,\ToKtag)$} A chaincode evaluates the predicate~$P$ and returns verdict~$v\in \{0,1\}$.
We consider as a sufficient proof the endorsement collection about the latter evaluation, i.e., $\ToKproof$ is a collection of signatures for a quorum of peers in the source chain. The tag~$\ToKtag$ is included in the signed statement, to ensure freshness of the information.

\paragraph*{$\ToKVerifyProof(\ToKproof,\ToKtag)$} Under our assumptions that the verifying client is always able to recompute the correct quorum of a given chain, the verification of $\ToKproof$ is simply a verification that the $\ToKtag$ is correct, that the signatures are valid and that the set of signers forms a valid quorum. Proofs can be verified by any device (e.g. computer/mobile device) or a blockchain (through a chaincode).

We implemented the transfer of asset as detailed below.

\paragraph{$\ToALock(a,A_t,C_t)$} The locking mechanism may vary with each use case, and it may follow different logics for each asset. In our evaluation setup, we added a \texttt{locked} field to the properties of our assets. When locking an asset, \texttt{locked} is set to~$\textsc{True}$ and the chaincode prevents any additional modification of the asset corresponding asset. The locking mechanism further includes a target chain~$C_t$ and an address~$A_t$ in order to prevent double spending. The proof is built as a proof of knowledge that the asset has indeed been locked.

\paragraph{$\ToAClaim(\ToAprooflock)$} An asset can be claimed on the destination chain as long as the tag~$\ToKtag$ of the knowledge proof is correct, in which case the asset is automatically transferred to the intended address~$A_t$. If the tag is incorrect, or the address is invalid, then the transaction fails ``successfully'', the failure is recorded to the chain, and a proof~$\ToAproofabort$ can be retrieved. Lastly, if the transaction is successful, the asset is transferred to the target address~$A_t$. A success transaction $\ToAproofclaim$ can then be generated to finalize the transfer on the original chain.

\paragraph{$\mathsf{ChainDivision}$}
We implemented the chain-division process in Fabric as a two-step process: first we create a new chain with the same state of the parent chain, then we assign the different peers between the two chain according to the assignment scheme in Section~\ref{sec:division:validator:assignment}.

\section{Implementation and Evaluation}
\label{sec:evaluation}

In this section, we evaluate the performance of our~\our{} based on Hyperledger Fabric.

\subsection{Implementation Setup}

We initialize a Fabric network with~$n$ peers and $n$~orderers inside docker containers, for up to~$44$ nodes, and we then proceed with a chain split resulting in~$\frac{n}{2}$ peers for each sibling chain.
Since the underlying consensus, Raft, is a CFT protocol, the number of faults in each sibling chain must be below~$50\%$ (i.e., $\alpha_i = \frac{1}{2}$ in our analysis, cf.~Figure~\ref{fig:plots:probability:sec:violation:50}). Recall that, as shown in Figure~\ref{fig:plots:probability:sec:violation:50}, \our{} preserves security with probability below~0.05 for~$n = 44$ as long as the faulty ratio~$\beta$ in the parent chain is below~$40\%$, i.e., as long as up to~$f = 17$ peers are faulty.

In each chain, we install the two following chaincodes:
a chain manager, for verifying proofs and monitoring the current chain configuration,
and an asset manager, which is responsible for cross-chain transfers.
We evaluate the time required for dividing chains, as well as for generating and verifying cross-chain transfers, for both asset and knowledge transfers.

For each experiment, we deploy the different docker containers on one server equipped with 6-Core Intel Xeon E-2176G and~128 GB RAM.
We consider different configurations by varying the number of nodes, for~$2\leq n \leq 44$, and we measure the performance of the various operations for each configuration.
In the case of transfer of knowledge, we measure the performance of the individual operations (i.e., proof generation and proof verification).
As for the performance of asset transfer, we measure the overall latency for an end-user to transfer an asset from one chain another.
Finally, we evaluate chain division by measuring the total downtime 
caused by this operation.

\subsection{Evaluation Results}

The results of our evaluation are shown in Figure~\ref{fig:evaluation} (each datapoint averages the results of~at least 5 independent runs).
Every plot illustrates the performance trend of a given operation as the number of peers~$n$ increases.
More specifically, Figure~\ref{fig:proofsize} shows the size of a proof of knowledge while Figure~\ref{fig:proofgentime} depicts the time required to generate the proof.
Figures~\ref{fig:proofverifoff} and~\ref{fig:proofverifon} show the time required to verify a proof, off-chain (i.e. locally) and on-chain (i.e. by a smart contract), respectively.
Finally, Figures~\ref{fig:ccsize} and~\ref{fig:cclatency} show proof size and latency required to perform a cross-chain transfer, compared to regular in-chain transfers.

\vspace{1 em}\noindent\textbf{Impact of chain splitting.}
The chain-division operation triggers a complex process that requires running a validator assignment scheme and a reconfiguration process to set up the two sibling chains.
In additional experiments, we measured the latency of the chain division process in terms of incurred downtime while the division process is ongoing.
Our experiments show that the chain-splitting latency grows linearly with the number of peers in the parent chain, e.g., it is about~$35$~seconds for a parent chain containing~$n=10$ peers, and about~$72$ seconds for~$n=20$, which is negligible compared to the frequency of chain splitting, which may range between days and months.
The linear dependency can be partly due to the bootstrapping of each container, which approximatively takes constant time.
We argue that the downtime incurred is manageable, especially because division happens rarely.

\vspace{1 em}\noindent\textbf{Performance of Transfer of Knowledge.}
In Figure~\ref{fig:proofsize}, we analyze the size of a proof of knowledge.
Since a proof consists of the concatenation of the various peers' signatures, its size grows linearly with the number of nodes.
An optimized implementation would aggregate the signatures rather than simply concatenating them, allowing to go from linear to constant size. This is a limitation of the current Fabric implementation---which does not allow aggregation.
Notice that the transaction inclusion proof is roughly twice as big as the generic proof: indeed, the inclusion proof contains i) a fully endorsed original transaction, and ii) an endorsement that such transaction has been executed, thereby combining endorsement collections twice.
Finally, the latency to generate and to verify a proof, respectively, are shown in Figures~\ref{fig:proofgentime}--\ref{fig:proofverifon}.
We observe a similar trend in all cases: the latency grows linearly with the number of nodes, again because of the non-aggregated signatures to be collected, resp. verified, for all peers.
Besides, notice that verifying an off-chain proof (Figure~\ref{fig:proofverifoff}) is much faster than an on-chain verification (Figure~\ref{fig:proofverifon}), on average by a factor of~4, due to the amount of endorsements to be collected for on-chain transactions in Fabric.
We emphasize that these costs are Fabric-specific, and hence are shared among all cross-chain transfer implementations.

\vspace{1 em}\noindent\textbf{Performance of Cross-chain transfer.} Figures~\ref{fig:ccsize} shows the average size for a cross-chain asset transfer, combining~3 messages for lock, claim, and validate operations, and illustrates also the size of an in-chain transaction (that changes the ownership of an asset within the same chain) for comparison.
Similarly, Figure~\ref{fig:cclatency} reports the latency of a cross-chain asset transfer, measured as the time interval starting with the generation of a lock transaction until the corresponding validate transaction has been fully ordered.
Again, both size and latency scale linearly with the number of peers, being the asset transfer based on three knowledge transfers (i.e., the same arguments as above apply).

\section{Conclusion}
\label{sec:conclusion}

Scalability remains one of the major challenges that hinders the adoption of permissioned blockchains in real-world applications. While the literature features a number of contributions that propose the reliance on sharding within permissioned blockchains, all existing solutions make the implicit assumption that all participating nodes are fixed upfront and keep participating in the consensus throughout the lifetime of the system.

In this paper, we presented, \our, the first solution for permissioned blockchains that supports the dynamic construction of shards, allows nodes to join and leave shards at will, and enables heterogenous shards to form and interact.
Inspired by cell mitosis, \our{} triggers sharding under high participation, and  merges shards in case of low participation. As far as we are aware, \our{} emerges as the first solution for permissioned blockchains that allows nodes to reactively self-organize to meet optimal performance.

We analyzed the security of \our{} and showed that, under mild assumptions on the number of faults among participants, our proposal to dynamically create shards via chain-division does not compromise the security of the blockchain ecosystem. We also implemented \our{} and integrated it within Hyperledger Fabric. Our evaluation results show that~\our{} incurs little modifications and negligible overhead when integrated with Hyperledger Fabric.

\section*{Acknowledgments}
The authors would like to thank the anonymous reviewers for their constructive feedback. This work was supported in part by the by the European Commission H2020 TeraFlow Project under Grant Agreement No 101015857.

\bibliographystyle{IEEEtranS}
\bibliography{IEEEabrv,references}

\begin{thebibliography}{10}
\providecommand{\url}[1]{#1}
\csname url@samestyle\endcsname
\providecommand{\newblock}{\relax}
\providecommand{\bibinfo}[2]{#2}
\providecommand{\BIBentrySTDinterwordspacing}{\spaceskip=0pt\relax}
\providecommand{\BIBentryALTinterwordstretchfactor}{4}
\providecommand{\BIBentryALTinterwordspacing}{\spaceskip=\fontdimen2\font plus
\BIBentryALTinterwordstretchfactor\fontdimen3\font minus
  \fontdimen4\font\relax}
\providecommand{\BIBforeignlanguage}[2]{{%
\expandafter\ifx\csname l@#1\endcsname\relax
\typeout{** WARNING: IEEEtranS.bst: No hyphenation pattern has been}%
\typeout{** loaded for the language `#1'. Using the pattern for}%
\typeout{** the default language instead.}%
\else
\language=\csname l@#1\endcsname
\fi
#2}}
\providecommand{\BIBdecl}{\relax}
\BIBdecl

\bibitem{DBLP:conf/sigmod/AmiriAA21}
M.~J. Amiri, D.~Agrawal, and A.~E. Abbadi, ``Sharper: Sharding permissioned
  blockchains over network clusters,'' in \emph{{SIGMOD} Conference}.\hskip 1em
  plus 0.5em minus 0.4em\relax {ACM}, 2021, pp. 76--88.

\bibitem{DBLP:conf/esorics/AndroulakiCCK18}
E.~Androulaki, C.~Cachin, A.~D. Caro, and E.~Kokoris{-}Kogias, ``Channels:
  Horizontal scaling and confidentiality on permissioned blockchains,'' in
  \emph{{ESORICS} {(1)}}, ser. Lecture Notes in Computer Science, vol.
  11098.\hskip 1em plus 0.5em minus 0.4em\relax Springer, 2018, pp. 111--131.

\bibitem{DBLP:Avarikioti:etal:2020}
\BIBentryALTinterwordspacing
G.~Avarikioti, E.~Kokoris-Kogias, and R.~Wattenhofer, ``Divide and scale:
  Formalization of distributed ledger sharding protocols,'' \emph{CoRR}, vol.
  abs/1910.10434, 2019. [Online]. Available:
  \url{http://arxiv.org/abs/1910.10434}
\BIBentrySTDinterwordspacing

\bibitem{peggedsidechains}
A.~Back, M.~Corallo, L.~Dashjr, M.~Friedenbach, G.~Maxwell, A.~Miller,
  A.~Poelstra, J.~Timón, and P.~Wuille, ``Enabling blockchain innovations with
  pegged sidechains,'' \url{https://blockstream.com/sidechains.pdf}, 2014.

\bibitem{DBLP:conf/osdi/CastroL99}
M.~Castro and B.~Liskov, ``Practical byzantine fault tolerance,'' in
  \emph{{OSDI}}.\hskip 1em plus 0.5em minus 0.4em\relax {USENIX} Association,
  1999, pp. 173--186.

\bibitem{DBLP:journals/corr/abs-2103-00254}
D.~Chaum, C.~Grothoff, and T.~Moser, ``How to issue a central bank digital
  currency,'' \emph{CoRR}, vol. abs/2103.00254, 2021.

\bibitem{DBLP:conf/sigmod/DangDLCLO19}
H.~Dang, T.~T.~A. Dinh, D.~Loghin, E.~Chang, Q.~Lin, and B.~C. Ooi, ``Towards
  scaling blockchain systems via sharding,'' in \emph{{SIGMOD}
  Conference}.\hskip 1em plus 0.5em minus 0.4em\relax {ACM}, 2019, pp.
  123--140.

\bibitem{gearbox}
B.~David, B.~Magri, C.~Matt, J.~B. Nielsen, and D.~Tschudi, ``Gearbox: An
  efficient {UC} sharded ledger leveraging the safety-liveness dichotomy,''
  \emph{{IACR} Cryptol. ePrint Arch.}, p. 211, 2021.

\bibitem{DBLP:conf/ccs/GervaisKWGRC16}
A.~Gervais, G.~O. Karame, K.~W{\"{u}}st, V.~Glykantzis, H.~Ritzdorf, and
  S.~Capkun, ``On the security and performance of proof of work blockchains,''
  in \emph{{CCS}}.\hskip 1em plus 0.5em minus 0.4em\relax {ACM}, 2016, pp.
  3--16.

\bibitem{DBLP:journals/iacr/Kokoris-KogiasJ17}
E.~Kokoris{-}Kogias, P.~Jovanovic, L.~Gasser, N.~Gailly, and B.~Ford,
  ``Omniledger: {A} secure, scale-out, decentralized ledger,'' \emph{{IACR}
  Cryptol. ePrint Arch.}, vol. 2017, p. 406, 2017.

\bibitem{BCC:LSFK2017}
W.~Li, A.~Sforzin, S.~Fedorov, and G.~Karame, ``{Towards Scalable and Private
  Industrial Blockchains},'' in \emph{Proceedings of the ACM Workshop on
  Blockchain, Cryptocurrencies and Contracts}.\hskip 1em plus 0.5em minus
  0.4em\relax {ACM}, 2017, pp. 9--14.

\bibitem{FastBFT}
J.~Liu, W.~Li, G.~O. Karame, and N.~Asokan, ``Scalable byzantine consensus via
  hardware-assisted secret sharing,'' \emph{{IEEE} Trans. Computers}, vol.~68,
  no.~1, pp. 139--151, 2019.

\bibitem{DBLP:conf/ccs/LuuNZBGS16}
L.~Luu, V.~Narayanan, C.~Zheng, K.~Baweja, S.~Gilbert, and P.~Saxena, ``A
  secure sharding protocol for open blockchains,'' in \emph{{CCS}}.\hskip 1em
  plus 0.5em minus 0.4em\relax {ACM}, 2016, pp. 17--30.

\bibitem{bitcoin}
S.~Nakamoto, ``Bitcoin: A peer-to-peer electronic cash system,'' 2008.

\bibitem{DBLP:conf/usenix/OngaroO14}
D.~Ongaro and J.~K. Ousterhout, ``In search of an understandable consensus
  algorithm,'' in \emph{{USENIX} Annual Technical Conference}.\hskip 1em plus
  0.5em minus 0.4em\relax {USENIX} Association, 2014, pp. 305--319.

\bibitem{bitcoinlightning}
J.~Poon and T.~Dryja, ``The bitcoin lightning network: Scalable off-chain
  instant payments,''
  \url{https://lightning.network/lightning-network-paper.pdf}, 2016.

\bibitem{tangle}
S.~Popov, ``The tangle,''
  \url{https://assets.ctfassets.net/r1dr6vzfxhev/2t4uxvsIqk0EUau6g2sw0g/45eae33637ca92f85dd9f4a3a218e1ec/iota1_4_3.pdf},
  2018.

\bibitem{raiden}
``The raiden network,'' \url{https://raiden.network/}, accessed: 2021-06-28.

\bibitem{DLTswitzerland}
F.~C. report, ``Legal frameworkfor distributed ledger technology and blockchain
  in switzerland,'' 2018.

\bibitem{skala2013hypergeometric}
M.~Skala, ``{Hypergeometric tail inequalities: ending the insanity},'' 2013.

\bibitem{DBLP:journals/iacr/SompolinskyLZ16}
Y.~Sompolinsky, Y.~Lewenberg, and A.~Zohar, ``{SPECTRE:} {A} fast and scalable
  cryptocurrency protocol,'' \emph{{IACR} Cryptol. ePrint Arch.}, vol. 2016, p.
  1159, 2016.

\bibitem{DBLP:journals/iacr/SompolinskyZ13}
Y.~Sompolinsky and A.~Zohar, ``Accelerating bitcoin's transaction processing.
  fast money grows on trees, not chains,'' \emph{{IACR} Cryptol. ePrint Arch.},
  vol. 2013, p. 881, 2013.

\bibitem{DBLP:conf/nsdi/WangW19}
J.~Wang and H.~Wang, ``Monoxide: Scale out blockchains with asynchronous
  consensus zones,'' in \emph{{NSDI}}.\hskip 1em plus 0.5em minus 0.4em\relax
  {USENIX} Association, 2019, pp. 95--112.

\bibitem{Hotstuff}
M.~Yin, D.~Malkhi, M.~K. Reiter, G.~Golan{-}Gueta, and I.~Abraham, ``Hotstuff:
  {BFT} consensus with linearity and responsiveness,'' in \emph{{PODC}}.\hskip
  1em plus 0.5em minus 0.4em\relax {ACM}, 2019, pp. 347--356.

\bibitem{DBLP:conf/ccs/ZamaniM018}
M.~Zamani, M.~Movahedi, and M.~Raykova, ``Rapidchain: Scaling blockchain via
  full sharding,'' in \emph{{CCS}}.\hskip 1em plus 0.5em minus 0.4em\relax
  {ACM}, 2018, pp. 931--948.

\bibitem{cryptoeprint:2019:1128}
A.~Zamyatin, M.~Al-Bassam, D.~Zindros, E.~Kokoris-Kogias, P.~Moreno-Sanchez,
  A.~Kiayias, and W.~J. Knottenbelt, ``Sok: Communication across distributed
  ledgers,'' \url{https://eprint.iacr.org/2019/1128}, 2019, financial
  Cryptography and Data Security 2021 (to appear).

\end{thebibliography}

\end{document}